\documentclass[useAMS,usenatbib,pdflscape]{mn2e}
\usepackage{graphicx}
\usepackage{epstopdf}
\usepackage{amsmath}
\usepackage{amssymb}
\usepackage{rotating}
\usepackage{pdflscape}
\usepackage{appendix}
\usepackage{subfigure}
\usepackage{stfloats}

\title[]{Effect of radiation drag on the line-force-driven winds}
\author[Wang et al.]{Bei-Chuan Wang$^1$, Xiao-Hong Yang$^1$\thanks{Corresponding author 1: yangxh@cqu.edu.cn (XH)}, De-Fu Bu$^2$\thanks{Corresponding author 2: dfbu@shao.ac.cn (DF)}, and Shu-Su Huang$^1$  \\
$^1$ Department of Physics and Chongqing Key Laboratory for Strongly Coupled Physic, Chongqing University, Chongqing 400044, China \\
$^{2}$ Key Laboratory for Research in Galaxies and Cosmology, Shanghai Astronomical Observatory, Chinese Academy of Sciences,\\ 80 Nandan Road, Shanghai 200030, China}

\begin{document}

\pagerange{\pageref{firstpage}--\pageref{lastpage}} \pubyear{20**}

\maketitle

\label{firstpage}
\def\LSUN{\rm L_{\odot}}
\def\MSUN{\rm M_{\odot}}
\def\RSUN{\rm R_{\odot}}
\def\MSUNYR{\rm M_{\odot}\,yr^{-1}}
\def\MSUNS{\rm M_{\odot}\,s^{-1}}
\def\MDOT{\dot{M}}

\begin{abstract}
Ultra-fast outflows (UFOs) with mildly relativistic velocities are measured using the X-ray spectra of radio-quiet and -loud active galactic nuclei (AGNs). In general, UFOs are believed to be generated from the accretion disk around a black hole (BH). A line-force driving model is suggested to be the mechanism to drive UFOs from the accretion disk. In this paper, we use the non-hydrodynamic approach to examine the influences of radiation-drag effects on the line-force-driven winds generated from the accretion disk. We find that the radiation-drag effects can significantly weaken the line-force-driven winds. Compared with the case without the radiation-drag effects, when the radiation-drag effects are considered, the maximum speed of winds is reduced by $\sim$60\%--70\%, the mass outflow rate is reduced by $\sim$50\%--80\%, and the kinetic power is reduced by about an order of magnitude. The radiation-drag effects narrow the area where the winds are generated.
\end{abstract}


\section{INTRODUCTION}

Blueshifted absorption lines are frequently detected on the ultraviolet (UV) and X-ray spectra of luminous AGNs (Churchill et al. 1999; Narayanan et al. 2004; Hamann et al. 1997; Rodr\'{\i}guez Hidalgo et al. 2011; Tombesi et al. 2010, 2012; Gofford et al. 2013, 2015). For example, it is reported that 10\%--17\% of the CIV absorption lines on the AGN UV spectra are blueshifted at 5000--70,000 km s$^{-1}$ (Misawa et al. 2007), and the Fe XXV and Fe XXVI absorption lines on the XMM-Newton spectra are blueshifted at ~0.03--0.3 times the speed ($c$) of light for $>$35 \% of radio-quiet AGNs (Tombesi et al. 2011). This implies that fast outflows/winds are common in AGNs and some winds can reach mildly relativistic velocities.

The highly blueshifted Fe XXV and Fe XXVI absorption lines imply that the highly ionized absorbers move fast outward from their nuclei with mildly relativistic velocities. Such outflows are called UFOs (Tombesi et al. 2010, 2012). Tombesi et al. (2010, 2011) first define UFOs to describe the highly ionized absorbers with velocities higher than 10$^4$ km s$^{-1}$ and derive the basic properties of UFOs. The ionization parameter ($\xi$) and column density ($N_{\rm H}$) of UFOs are in the range log($\xi$/(erg s$^{-1}$ cm))$\sim$3--6 and log($N_{\rm H}$/(cm$^{-2}$))$\sim$22-24, respectively (Tombesi et al. 2011). The UFOs are observed not only in a significant fraction ($>35\%$) of radio-quiet AGNs (Tombesi et al. 2011) but also in a small sample of radio-loud AGNs (Tombesi et al. 2014). In general, the UFOs are estimated to be formed in the interval of $10^2$--$10^4$ Schwarzschild radius ($r_{\rm s}$) away from the center BH (Tombesi et al. 2012; Gofford et al. 2015). This implies that the UFOs may be the winds generated from the accretion disk around the center BH (e.g. Proga \& Kallman 2004; Fukumura et al. 2010; Nomura et al. 2016; Nomura \& Ohsuga 2017; Yang et al. 2021; Yang 2021).

Two kinds of models, i.e. a magnetic-driving model and a line-force driving model, are proposed to produce UFOs from the accretion disk. The mechanism of the magnetic-driven winds and jets, called the Blandford \& Payne mechanism, has been developed in previous works (e.g. Blandford \& Payne 1982; sakurai 1987; Lovelace et al. 1987; Cao \& Spruit 1994; Fukmura et al. 2014, 2018; Chen et al. 2021). Fukmura et al. (2018) have used a model of magnetic-driven winds to model the UFOs observed in PDS 456. In the magnetic-driving model, a large-scale magnetic field is required. However, the formation of the large-scale magnetic field is still an open issue (e.g. Cao 2011; Cao \& Spruit 2013). The magnetic structure is often taken as a free parameter in the model of magnetic-driven winds.

A line-force driving model is the other promising model to understand UFOs. For luminous AGNs, such as quasars, a geometrically thin and optically thick disk is often proposed to be around the center BH (Shakura \& Sunyaev 1973). The thin disk is luminous and irradiates gases above the thin disk. The photons emitted from the thin disk are scattered or absorbed by the gases and therefore the Compton-scattering force and the spectral line force are produced and exerted on the gases. For the highly ionized gases, the Compton-scattering force is the main radiation force. When the effective barrier of gravity and centrifugal force of gases is small, the Compton-scattering force may drive winds (e.g. Cao 2014; Yang et al. 2018; 2019). For the weakly ionized gases, the line force caused by the absorption of UV photons becomes an important radiation force to drive winds (Castor et al. 1975; Murray et al. 1995). The Compton-scattering force cannot drive the winds with mildly relativistic velocities. The line force cannot drive the highly ionized winds, because the line force is negligible for the gases with the ionization parameter higher than 100 erg s$^{-1}$ cm. For UFOs, their ionization parameter is much higher than 100 erg s$^{-1}$ cm. Yang et al. (2021) pointed out that UFOs can be intermittently accelerated by the line force. When the X-ray photons from the center corona are shielded by dense gases in the inner region, the degree of ionization of gases becomes so low that the line force can drive the gases to form winds. When the line-force driven winds are exposed to the X-ray photons, the winds become highly ionized and then form the observed UFOs. Therefore, the line-force driven winds may be the origin of UFOs (Nomura \& Ohsuga 2017; Mizumoto et al. 2021; Yang et al. 2021; Yang 2021)

For the radiation-driven winds, such as the line-force-driven winds, if they move at mildly-relativistic velocity in the radiation field, the special relativistic effects (radiation drag) become important. When the relativistic effects are taken into account, the wind speed is reduced up to 50\% with respect to the non-relativistic case (Luminari et al. 2020, 2021). Quera-Bofarull et al. (2021) further found that the line-force-driven winds is mildly relativistic, with velocity 0.1--0.3 c, when the relativistic effects are included. In Quera-Bofarull et al.' work (2021), they took into account a correction to the radiation flux seen by the high-velocity winds, while they ignored the influence of the disk rotation on the radiation field in the rest frame standing at the infinity.  Therefore, our motivation in this paper is to continue to study the influence of the radiation-drag effects on the line-force-driven winds.

In term of research methods, the line-force-driven winds in AGNs have been studied not only in radiation hydrodynamic simulations (e.g. Proga et al. 2000; Proga \& Kallman 2004; Nomura \& Ohsuga 2017; Yang et al. 2021; Yang 2021) but also in non-hydrodynamic approach (Risaliti \& Elvis 2010; Nomura et al. 2013). In the non-hydrodynamic approach, the gas pressure is neglected, gases are taken as particles, and then the Lagrangian method is employed to model the line-force-driven winds. Compared with the radiation hydrodynamic simulations, the non-hydrodynamic approach takes less time in calculation and then can test more parameter space. Therefore, in this paper, we employ the non-hydrodynamic approach to study the line-force-driven winds.

\begin{table*}
\begin{center}

\caption[]{Summary of Models}

\begin{tabular}{ccccccccccc}
\hline\noalign{\smallskip} \hline\noalign{\smallskip}
 Models & $L_{\rm bol}$  &  $v_0$ & $n_0$ &$v_{\rm max}$ & $\dot{M}_{\rm out}$ & $P_{\rm kin}$& $\theta_{\rm out}$\\
&$(L_{\rm Edd}) $& $ ({\rm m/s})$ & $ ({\rm cm^{-3}})$ &$(c)$ & $(M_{\odot} {\rm year}^{-1}$) & $(L_{\rm Edd}$) & $(^{\circ})$ \\

\hline\noalign{\smallskip}
A0  & $0.01$&$1 \times {10^5} $&$5 \times {10^7} $&$0.035$ & 0.003& $9.00\times {10^{-6}}$  &  9.4--10.4 \\
A1  & $0.05$&$1 \times {10^5} $&$5 \times {10^7} $&$0.176$ & 0.035& $9.43\times {10^{-4}}$  & 7.3--13.0  \\
A2  & $0.1$&$1 \times {10^5} $&$5 \times {10^7} $&$0.228$ & 0.069& $2.27\times {10^{-3}}$  & 5.8--12.6  \\
A3  & $0.2$&$1 \times {10^5} $&$5 \times {10^7} $&$0.288$ &0.113 & $4.58 \times {10^{-3}}$ & 4.1--12.1\\
A4  & $0.3$&$1 \times {10^5} $&$5 \times {10^7} $&$0.316$ &0.137 & $6.42 \times {10^{-3}}$ & 6.8--11.8\\
A5  & $0.4$&$1 \times {10^5} $&$5 \times {10^7} $&$0.317$&0.153 & $7.74 \times {10^{-3}}$  & 7.0--10.2\\
A6  & $0.5$&$1 \times {10^5} $&$5 \times {10^7} $&$0.331$&0.163 & $9.24 \times {10^{-3}}$  & 7.1--11.2\\
A7  & $0.6$&$1 \times {10^5} $&$5 \times {10^7} $&$0.353$&0.173& $1.00 \times {10^{-2}}$  & 7.3--10.8\\
A8  & $0.7$&$1 \times {10^5} $&$5 \times {10^7} $&$0.335$&0.190 & $1.10 \times {10^{-2}}$  & 7.5--10.9\\
A9  & $0.8$&$1 \times {10^5} $&$5 \times {10^7} $&$0.360$&0.183 & $1.15 \times {10^{-2}}$  & 7.6--10.4\\

A10  & $0.2$&$1 \times {10^4} $&$5 \times {10^7} $&$0.384$&0.019 & $8.89 \times {10^{-4}}$  &7.6--14.6\\
A11  & $0.2$&$5 \times {10^4} $&$5 \times {10^7} $&$0.317$&0.056 & $2.96 \times {10^{-3}}$  &4.6--12.3\\
A12  & $0.2$&$5 \times {10^5} $&$5 \times {10^7} $&$0.203$&0.435 & $9.40 \times {10^{-3}}$  &6.6--11.7\\
A13  & $0.2$&$1 \times {10^6} $&$5 \times {10^7} $&$0.170$&0.870 & $1.42 \times {10^{-2}}$  &10.1--11.9\\

A14  & $0.2$&$1 \times {10^5} $&$1 \times {10^7} $&$0$&0 & $0$ &0 \\
A15  & $0.2$&$1 \times {10^5} $&$1 \times {10^8} $&$0.437$&0.112 & $1.02\times {10^{-2}}$ & 6.7--12.9\\
A16  & $0.2$&$1 \times {10^5} $&$5 \times {10^8} $&$0.621$&0.046 & $2.18 \times {10^{-2}}$ & 14.7--18.2\\
A17  & $0.2$&$1 \times {10^5} $&$1 \times {10^9} $&$0.909$&0.063 & $5.02 \times {10^{-2}}$ & 16.2--22.6\\
\hline\noalign{\smallskip}
B0  & $0.01$&$1 \times {10^5} $&$5 \times {10^7} $&$0$ & 0& $0$  &  0\\
B1  & $0.05$&$1 \times {10^5} $&$5 \times {10^7} $&$0$ & 0& $0$  & 0\\
B2  & $0.1$&$1 \times {10^5} $&$5 \times {10^7} $&$0.08$&0.012&$1.08 \times {10^{-4}}$ & 7.4--9.2 \\
B3  & $0.2$&$1 \times {10^5} $&$5 \times {10^7} $&$0.112$&0.029&$4.36 \times {10^{-4}}$ & 6.7--9.6\\
B3a  & $0.2$&$1 \times {10^5} $&$5 \times {10^7} $&$0.118$&0.032&$4.72 \times {10^{-4}}$ &7.1--10.6\\

B4  & $0.3$&$1 \times {10^5} $&$5 \times {10^7} $&$0.129$&0.048&$7.26 \times {10^{-4}}$ & 6.2--9.9\\
B5  & $0.4$&$1 \times {10^5} $&$5 \times {10^7} $&$0.138$&0.063&$8.38 \times {10^{-4}}$ & 5.4--10.1\\
B6  & $0.5$&$1 \times {10^5} $&$5 \times {10^7} $&$0.136$&0.068&$1.16 \times {10^{-3}}$ & 5.2--9.3\\
B7  & $0.6$&$1 \times {10^5} $&$5 \times {10^7} $&$0.150$&0.084&$1.44 \times {10^{-3}}$ & 4.5--8.6\\
B8  & $0.7$&$1 \times {10^5} $&$5 \times {10^7} $&$0.144$&0.087&$1.46 \times {10^{-3}}$ & 4.1--9.0\\
B9  & $0.8$&$1 \times {10^5} $&$5 \times {10^7} $&$0.152$&0.087&$1.57 \times {10^{-3}}$ & 4.4--9.6\\

B10  & $0.2$&$1 \times {10^4} $&$5 \times {10^7} $&$0.202$&0.010 & $2.88 \times {10^{-4}}$ & 4.9--9.8\\
B11  & $0.2$&$5 \times {10^4} $&$5 \times {10^7} $&$0.138$&0.024 & $4.67 \times {10^{-4}}$ & 5.9--10.1\\
B12  & $0.2$&$5 \times {10^5} $&$5 \times {10^7} $&$0.042$&0.031 & $1.25 \times {10^{-4}}$ & 7.9--8.7\\
B13  & $0.2$&$1 \times {10^6} $&$5 \times {10^7} $&$0$&0 & $0$ & 0\\

B14  & $0.2$&$1 \times {10^5} $&$1 \times {10^7} $&$0$&0& $0$ & 0\\
B15  & $0.2$&$1 \times {10^5} $&$1 \times {10^8} $&$0.190$&0.028 & $1.19 \times {10^{-3}}$ & 9.1--12\\
B16  & $0.2$&$1 \times {10^5} $&$5 \times {10^8} $&$0.224$&0.019 & $2.18 \times {10^{-3}}$ & 12.1--20.4\\
B17  & $0.2$&$1 \times {10^5} $&$1 \times {10^9} $&$0.314$&0.024 & $5.40 \times {10^{-3}}$ & 17.3--21.8\\

\hline\noalign{\smallskip}
\hline\noalign{\smallskip}
\end{tabular}
\end{center}
\begin{list}{}
\item\scriptsize{\textit{Note}. Column (1): the model names; Column (2): the bolometric luminosity; Column (3): the vertical velocity at the streamline foot; Column (4): the initial particle number density of streamlines; Column (5): the fastest velocity along streamlines; Columns (6) and (7): the mass outflow rate and kinetic power of the escape particles, Column (8): the range of the velocity azimuth angles (relative to the disk surface) of the escape particles at the outer boundary, respectively. We set the black hole mass to be $10^8 M_{\odot}$, and the rate ($f_X$) of the X-ray luminosity to the bolometric luminosity to be 0.1, respectively. Models A0--A17 donot include the radiation-drag effects, while models B0--B17 include the radiation-drag effects. For models A14, B0, B1, B13, and B14, their $v_{\rm max}$, $\dot{M}_{\rm out}$, and $P_{\rm kin}$ equal 0, which means that the escape particles are not generated.}
\end{list}

\end{table*}

\begin{figure*}
\scalebox{0.45}[0.45]{\rotatebox{0}{\includegraphics{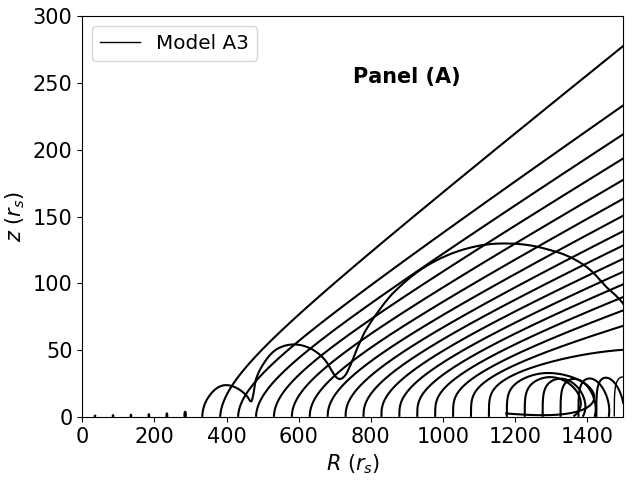}}}
\scalebox{0.45}[0.45]{\rotatebox{0}{\includegraphics{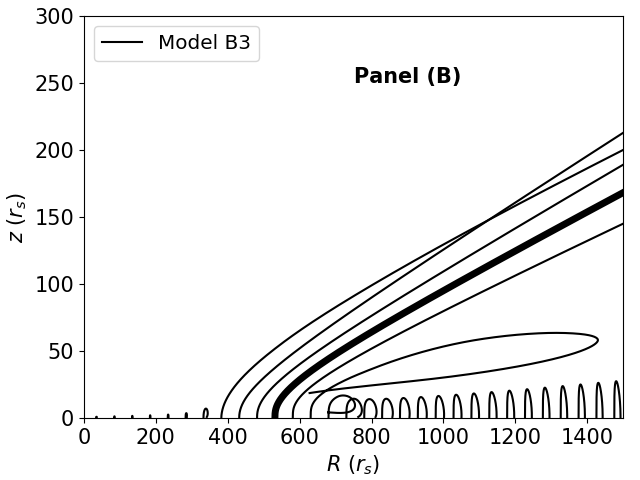}}}
\scalebox{0.45}[0.45]{\rotatebox{0}{\includegraphics{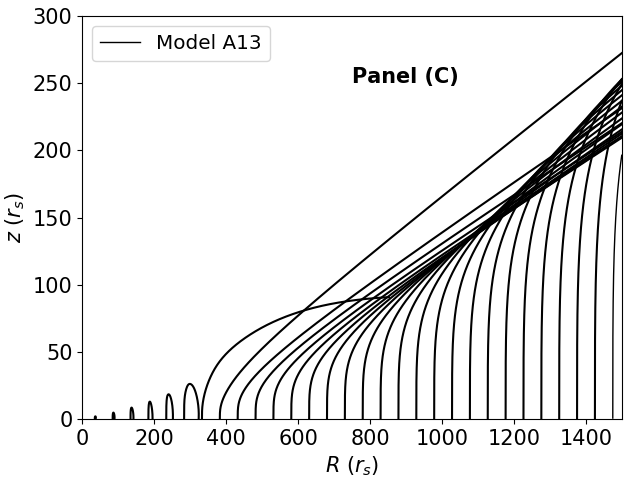}}}
\scalebox{0.45}[0.45]{\rotatebox{0}{\includegraphics{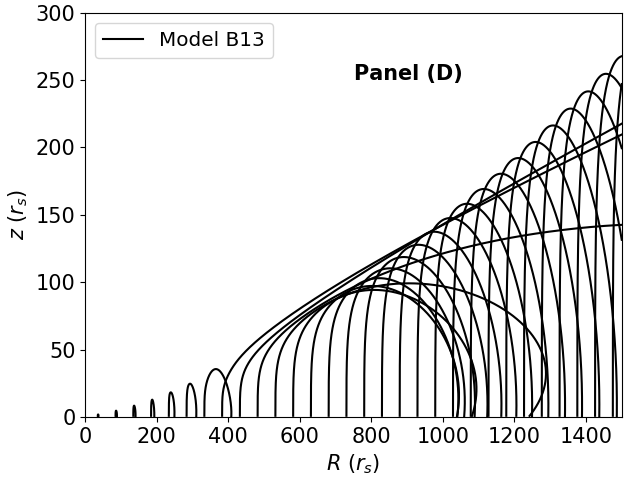}}}
\ \centering \caption{The trajectories of winds in models A3 (panel (A)), B3 (panel (B)), A13 (panel (C)), and B13 (panel (D)). }\label{Figure 1.}
\end{figure*}

\begin{figure*}
\scalebox{0.5}[0.5]{\rotatebox{0}{\includegraphics{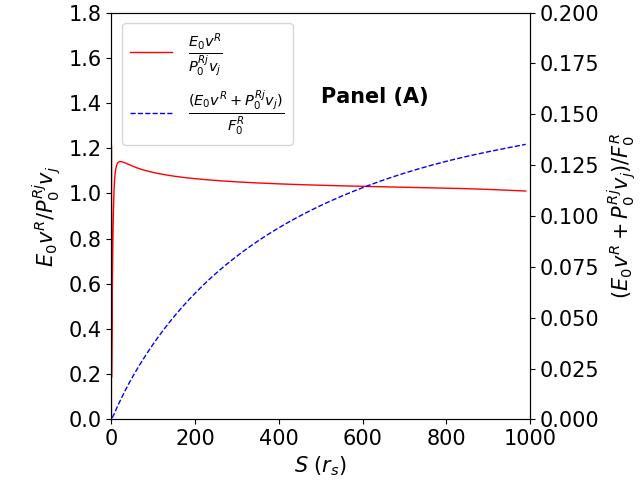}}}
\scalebox{0.5}[0.5]{\rotatebox{0}{\includegraphics{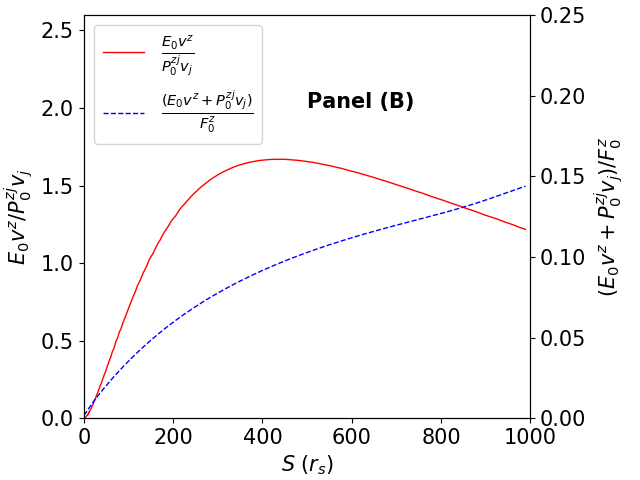}}}
\scalebox{0.5}[0.5]{\rotatebox{0}{\includegraphics{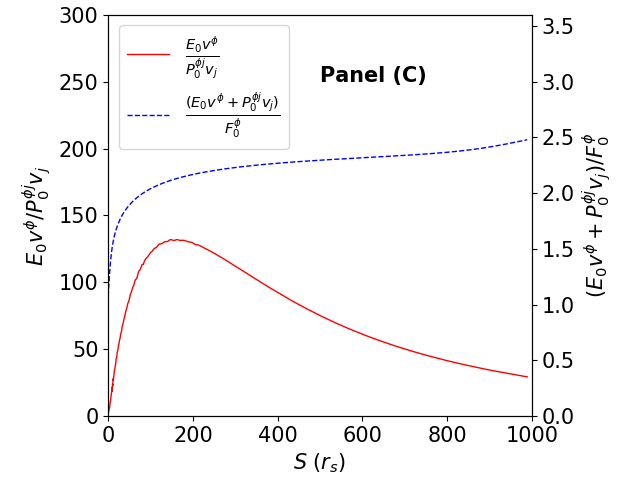}}}
\ \centering \caption{ Comparison of terms in equation (7). Panel (A) shows the $(E_0v^R+ v_j P_0^{Rj})/{F^{R}_0}$ and ${E_0v^R}/{v_j P_0^{Rj}}$ values along the thick line in panel (B) of figure 1. Panel (B) shows the $(E_0v^z+v_j P_0^{zj})/{F^{z}_0}$ and ${E_0v^z}/{v_j P_0^{zj}}$ values along the thick line in panel (B) of figure 1. Panel (C) shows the $(E_0v^\phi+v_j P_0^{\phi j})/{F^{\phi}_0}$ and ${E_0v^\phi}/{v_j P_0^{\phi j}}$ values along the thick line in panel (B) of figure 1.}\label{Figure 2.}
\end{figure*}

\begin{figure*}
\scalebox{0.45}[0.45]{\rotatebox{0}{\includegraphics{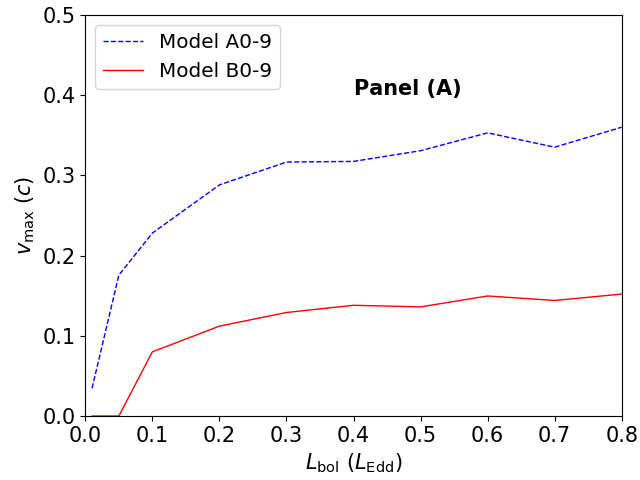}}}
\scalebox{0.45}[0.45]{\rotatebox{0}{\includegraphics{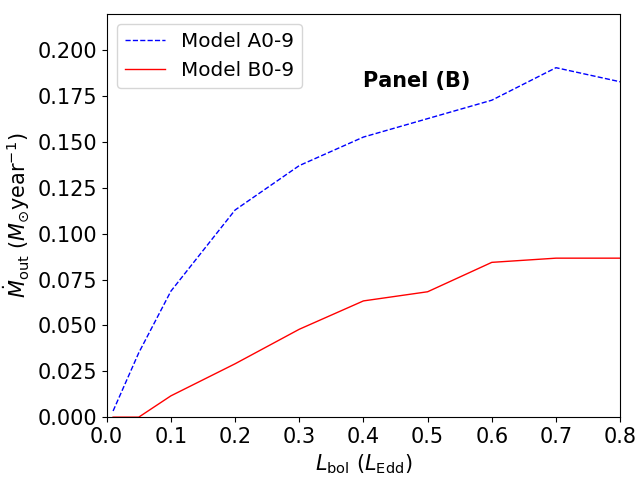}}}
\scalebox{0.45}[0.45]{\rotatebox{0}{\includegraphics{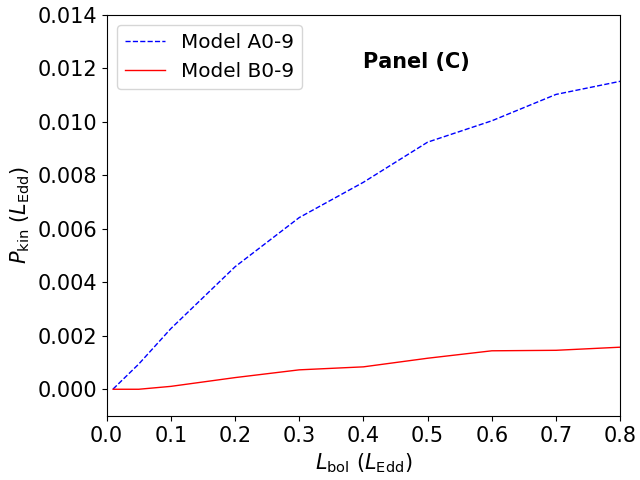}}}
\scalebox{0.45}[0.45]{\rotatebox{0}{\includegraphics{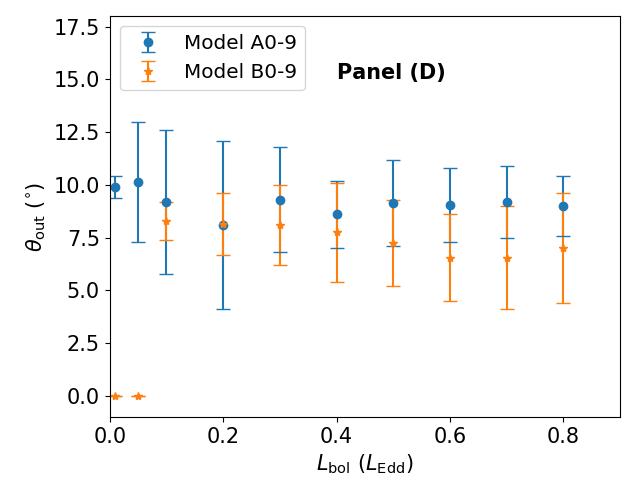}}}
\ \centering \caption{ The influences of the bolometric luminosity ($L_{\rm bol}$) on the wind properties, such as the maximum speed ($v_{\rm max}$) along streamlines (Panel (A)), the mass outflow rate ($\dot M_{\rm out}$, Panel (B)), the kinetic power ($P_{\rm kin}$, Panel (C)), the angle of winds ($\theta_{\rm out}$, Panel (D)).}\label{Figure 3.}
\end{figure*}

\begin{figure*}
\scalebox{0.52}[0.45]{\rotatebox{0}{\includegraphics{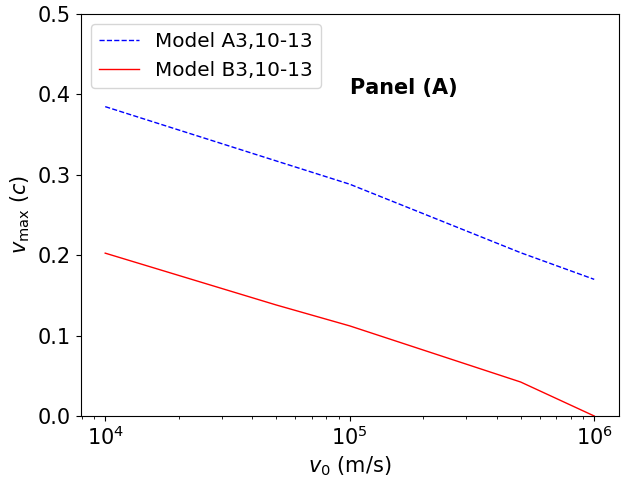}}}
\scalebox{0.52}[0.45]{\rotatebox{0}{\includegraphics{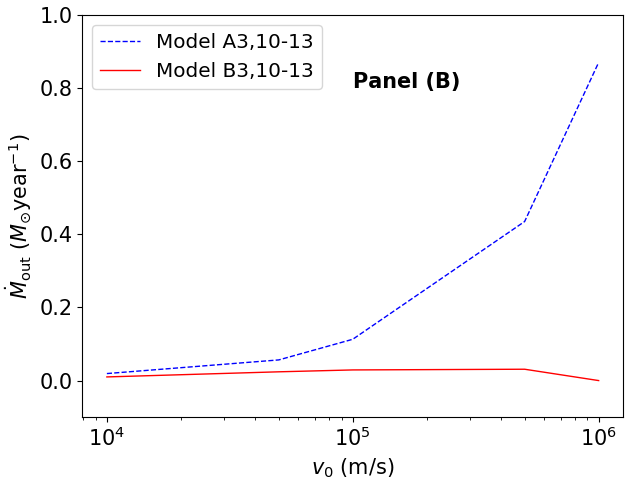}}}
\scalebox{0.52}[0.45]{\rotatebox{0}{\includegraphics{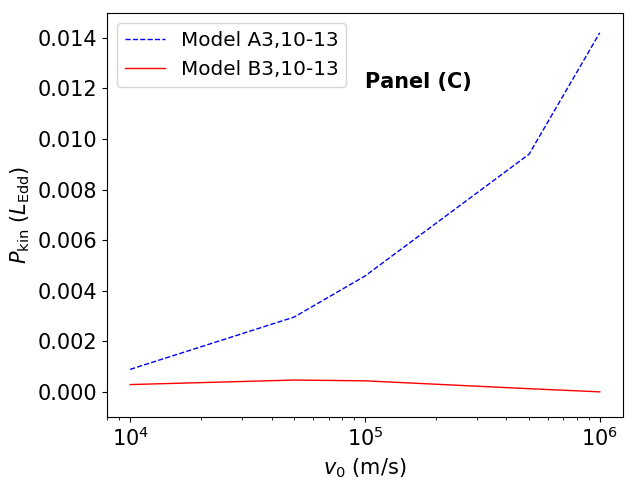}}}
\scalebox{0.52}[0.45]{\rotatebox{0}{\includegraphics{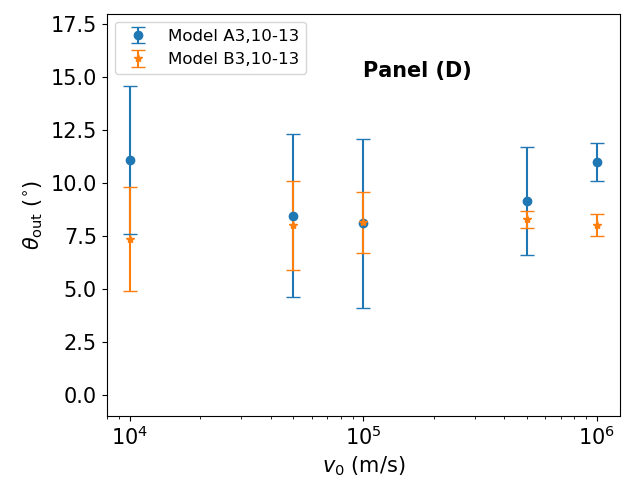}}}
\ \centering \caption{The influences of the initial velocity ($v_{0}$) on the wind properties. Panels (A)--(D) are the same as the panels of Figure 3.}\label{Figure 4.}
\end{figure*}

\begin{figure*}
\scalebox{0.45}[0.45]{\rotatebox{0}{\includegraphics{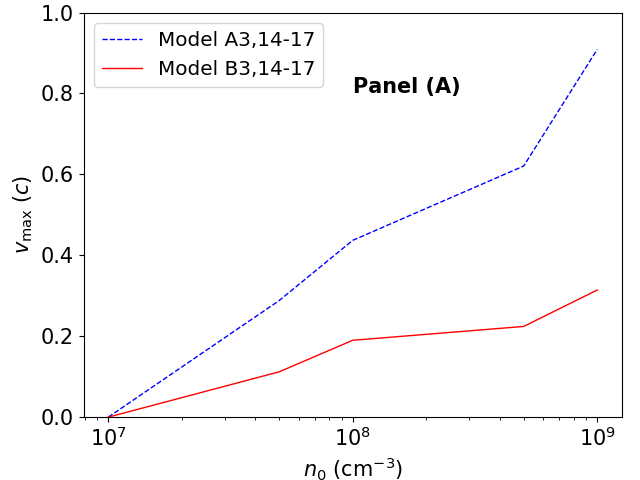}}}
\scalebox{0.45}[0.45]{\rotatebox{0}{\includegraphics{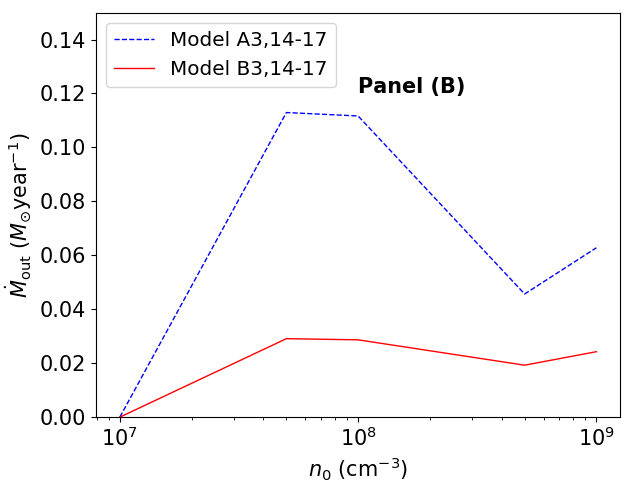}}}
\scalebox{0.45}[0.45]{\rotatebox{0}{\includegraphics{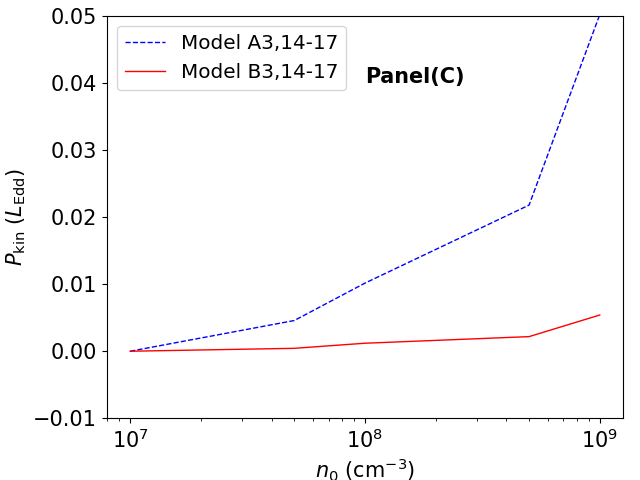}}}
\scalebox{0.45}[0.45]{\rotatebox{0}{\includegraphics{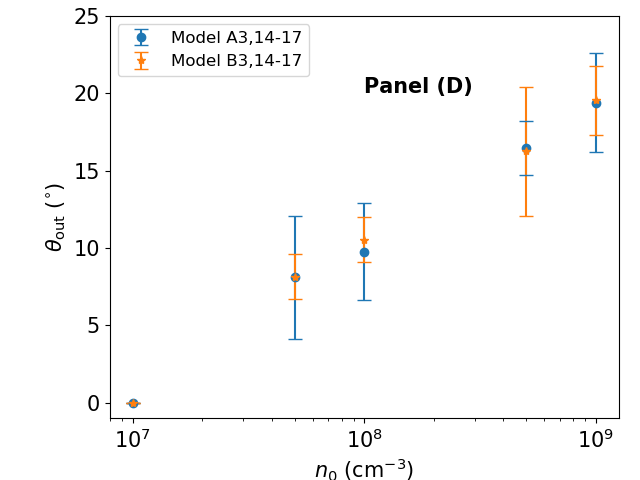}}}
\ \centering \caption{The influences of the number density ($n_{0}$) of particles on the wind properties. Panels (A)--(D) are the same as the panels of Figure 3.}\label{Figure 5.}
\end{figure*}

The paper is organized as follows. In section 2, we describe our model and method. In section 3, we present our results and related discussions. In section 4, we give conclusions and discussions.

\section{MODEL AND METHOD}

It is believed that a supermassive BH locates at the center of galaxies. In luminous AGNs, such as quasars, it is often believed that a standard thin disk exists around the supermassive BH. The standard thin disk is geometrically thin and optically thick (Shakura \& Sunyaev 1973). Observations imply that there is a hot corona within 10 Schwarzschild radius ($r_{\rm s}$) (Reis \& Miller 2013; Uttley et al. 2014). The gases above the thin disk are irradiated by the UV photons generated in the thin disk and the X-ray photons generated in the hot corona, respectively. We trace trajectories of the irradiated gases in the cylindrical coordinates ($R$,$\phi$,$z$). The trajectory lines are used to model the winds driven by the radiation emitted by the thin disk. For simplicity, we assume that the scale height of the thin disk is constant with the change in radius, set the disk surface to be the plane of $z=0$, and set the hot corona to locate at the center (i.e. $R=z=0$).

For the line-force-driven winds, their kinetic energy is much stronger than their thermal energy, which causes that the gas pressure is negligible in dynamics. Therefore, the line-force-driven winds are taken as particles. We can calculate the trajectory of a particle by solving the dynamic equations of the particles as follows:
\begin{equation}
\frac{dR}{dt}=v_{\textit{\tiny{R}}},
\end{equation}
\begin{equation}
\frac{d\phi}{dt}=\frac{1}{R}{v_{\phi}},
\end{equation}
\begin{equation}
\frac{dz}{dt}=v_{\textit{\tiny{z}}},
\end{equation}
\begin{equation}
\frac{dv_{\textit{\tiny{R}}}}{dt}=-\frac{G M_{\rm BH}R}{(R^2+z^2)^{3/2}}+\frac{v_{\phi}^2}{R}+\frac{\sigma_{\rm T}}{m_{\rm p} c}(1+\mathcal{M})F^{\textit{\tiny{R}}}e^{-\tau_{uv}},
\end{equation}

\begin{equation}
\frac{1}{R}\frac{d(R v_{\phi})}{dt}=\frac{\sigma_{\rm T}}{m_{\rm p} c}(1+\mathcal{M})F^{\phi}e^{-\tau_{uv}},
\end{equation}

\begin{equation}
\frac{dv_z}{dt}=-\frac{G M_{\rm BH}z}{(R^2+z^2)^{3/2}}+\frac{\sigma_{\rm T}}{m_{\rm p} c}(1+\mathcal{M})F^{z}e^{-\tau_{uv}},
\end{equation}
where $v_{i}\equiv(v_{\textit{\tiny{R}}}, v_{\phi}, \text{ and } v_{z})$ is the velocity of particles, $F^{i}\equiv(F^{\textit{\tiny{R}}}, F^{\phi}, \text{ and } F^{z})$ is the radiative flux measured in the comoving frame with the moving particles, $\mathcal{M}$ is the line-force multiplier, $\tau_{\rm uv}$ is the ultraviolet optical depth, and $\sigma_{\rm T}$, $m_{\rm p}$, $G$, and $M_{\rm BH}$ are the Thomson-scattering cross section, the proton mass,  the gravity constant, and the BH mass, respectively. The calculation of $\mathcal{M}$ is referred to Arnau Quera-Bofarull et al. (2020), we will discuss it in Appendix A. In equations (4)--(6), we ignore the gravitational force of the thin disk and the Compton-scattering force of the X-ray photons generated in the hot corona. The radiation field is divided into two components, i.e the X-ray radiation generated in the hot corona and the UV radiation generated in the thin disk. The UV radiation force is an important force of driving gases to form winds by the Compton-scattering force and the spectral line force. The X-ray radiation flux from the hot corona is much smaller than the UV radiation flux from the thin disk. The X-ray radiation is negligible in dynamics, while the X-ray photons play an important role in ionizing gases. We set the disk luminosity ($L_{\rm disk}$) to be $L_{\rm disk}=f_{\rm disk}L_{\rm bol}$ and the X-ray luminosity to be $L_{\rm X}=f_{\rm X}L_{\rm bol}=(1-f_{\rm disk})L_{\rm bol}$, where $L_{\rm bol}$ is the bolometric luminosity. To study the effect of radiation drag on the line-force-driven winds, the special relativistic effects is also included when the radiation flux from the thin disk is calculated. We have two different frame: the comoving frame with the moving particles and the rest frame standing at the infinity. In equations (4)--(6), $F^{i}$ is measured in the comoving frame. The radiation field generated in the thin disk is transferred into the comoving frame. On the order of $v/c$, the $i$th component of radiative flux in the comoving frame is given by
\begin{equation}
F^{i}=F^{i}_{0}-E_{0}v^{i}- v_{j}P^{ij}_{0},
\end{equation}
where $F^{i}_{0}$, $E_{0}$, and $P^{ij}_{0}$ are the $i$th component of radiative flux, the radiation energy density, and the radiation stress tensor in the rest frame, respectively. In equation (7), $v^{i}\equiv v_{i}$ (i.e. $v^{R}=v_{R}$, $v^{z}=v_{z}$, and $v^{\phi}=v_{\phi}$) and the Einstein summation convention is applied to $v_{j}P^{ij}_{0}$. In the previous works about the line-force-driven winds, $F^{i}_{0}$ in the rest frame does not include the Doppler effect of rotating of the disk, while in our calculation $F^{i}_{0}$, $E_{0}$, and $P^{ij}_{0}$ include the Doppler effect of rotating of the disk. The calculation of these quantities is referred to Tajima \& Fukue (1998). For the completeness of this paper, we also give the introduction of the calculation in Appendix B.

In equations (4)--(6), the line-force multiplier $\mathcal{M}$ is a function of the ionization parameter ($\xi$) and the local optical depth parameter ($t$) (Rybicki \& Hummer 1978), according to the Sobolev approximation. The ionization parameter $\xi$ is given by
\begin{equation}
\xi=\frac{L_{X}\text{exp}(-\tau_{X})}{n (R^2+z^2)},
\end{equation}
where $n$ is the number density and $\tau_{X}$ is the X-ray optical depth. According to Quera-Bofarull et al. (2020), $\tau_{X}$ can be given by
\begin{equation}
\tau_{X}\equiv\int_{R_{\rm b}}^{R_0}\sec(\theta) m_p n_0\sigma_X dR'+\int_{R_0}^R \sec(\theta) m_p n(r)\sigma_X dR',
\end{equation}
where $R_{\rm b}$ is the radius at which we start the first streamline, $R_0$ is the initial radius of the considered streamline, $\theta$ is the angle between the line from the particle to the center and the disk surface, $n_0$ is the particle number density on the disk surface, $n(r)$ is the particle number density located at $r=(R^2+z^2)^{\frac{1}{2}}$ in the considered streamline, and $\sigma_X$ is the X-ray attenuation, respectively. Following Proga et al. (2000), the X-ray attenuation is set to be $\sigma_X=0.4 \text{ g}^{-1}\text{cm}^2$ for $\xi\geq10^5$ erg s$^{-1}$ cm and $\sigma_X=40 \text{ g}^{-1}\text{cm}^2$ for $\xi<10^5$ erg s$^{-1}$ cm. Moreover, the ultraviolet optical depth $\sigma_{\rm uv}=0.4 \text{ g}^{-1}\text{cm}^2$ for any $\xi$. Similarly, the UV optical depth is given by
\begin{equation}
\tau_{uv}\equiv\int_{R_{\rm b}}^{R_0}\sec(\theta) m_p n_0\sigma_{\rm uv} dR'+\int_{R_0}^R \sec(\theta) m_p n(r)\sigma_{\rm uv} dR',
\end{equation}
and we set $\sigma_{\rm uv}=0.4 \text{ g}^{-1}\text{cm}^2$ for any $\xi$. Here, for simplicity, we employ the simply method to calculate the optical depth of UV and X-ray photons. In Quera-Bofarull et al.'s work (2021), the treatment of the optical depth is more accurate.

AGNs in our models have three parameters, i.e. the black hole mass ($M_{\rm BH}$), the bolometric luminosity ($L_{\rm bol}$), and the radio ($f_{\rm disk}$) of the disk luminosity to the bolometric luminosity. According to the disk luminosity, we can determine the accretion rate $\dot{M}=12 L_{\rm disk}/c^2$. Winds are launched from the disk surface. We use streamlines to describe winds. We simulate $30$ streamlines at equal intervals from 10 $r_{\rm s}$ to 1500 $r_{\rm s}$. For comparison, we employ 40 streamlines to calculate a model (model B3a in Table 1). We find that when the number of streamlines is more than 30, the results donot significantly change (see Table 1). The positions of streamline foot are set to be ($R_{0}$, $\phi_{0}=0$, $z_{0}$), where $R_{0}$ is the distance from the rotational axis and $z_{0}$ is the height from the disk surface, respectively. In the previous works, the disk height is often assumed to be constant with the change of radius for simplicity of calculation. We also assume the disk height to be constant and set $z_{0}=1$ $r_{\rm s}$. Then, the position of streamline foot is determined only by $R_{0}$. For calculating the streamlines of winds, besides $R_{0}$, the vertical velocity ($v_{0}$) and particle number density ($n_{0}$) of lunched winds need to be determined as the boundary conditions of streamlines. For a streamline, we assume its initial width to be $\Delta L_{0}$ and assume the streamline width $\Delta L\varpropto r$. According to the mass continuity equation, along a streamline the mass outflow rate ($\dot{M}_{\rm line}$) of winds is constant. Therefore, we have
\begin{equation}
\dot{M}_{\rm line}=m_{\rm p} n(r) v 2\pi r \Delta L = m_{\rm p} n_{0}v_{0}2\pi r_{0}\Delta L_0,
\end{equation}
where $r_{0}=R_{0}$ on the disk surface and $n(r)$ is the particle number density at $r=\sqrt{R^2+z^2}$. For simplicity, we assume $\Delta L\propto r$ along the streamlines. Then we have $\Delta L=\frac{r}{r_{0}} \Delta L_{0}$ and get $n(r)v r^2=n_{0}v_{0} r^2_{0}$. When the boundary conditions (i.e. $R_{0}$, $v_{0}$, and $n_0$) are given, we integrate equations (1)--(6), then get the position and velocity, and finally update the number density using equation (9). This process is iterated until streamline is obtained.

Our calculation is based on the QWIND code (Quera-Bofarull et al. 2020). In the QWIND code, the ASSIMULO simulation software package (Andersson et al. 2015) is employed and the IDA solver (Hindmarsh et al. 2004) is implemented to solve equations (1)--(6). The details of the QWIND code are referred to Quera-Bofarull et al. (2020). According to equation (7), we modified the QWIND code for considering the radiation-drag effects. This updated code, Dwind, is now available as a public release on GitHub.\footnote{https://github.com/wangbeicuan/Dwind.git}

\section{RESULTS}

\subsection{Effects of radiation drag.}

The radiation forces exerted on particles are determined by the radiation fluxes ($F^i$) measured in the comoving frame with the moving particles. According to equation (7),  the measured radiation fluxes ($F^i$) in the comoving frame include two terms: one is the radiation fluxes ($F^i_{0}$) in the rest frame standing at the infinity, and other term ($-E_0 v^i-v_j P^{ij}_0$) is the radiation-drag effect, which is attributed to the relativistic effect due to that particles move in radiation field at high speed. The radiation-drag force is proportional to the velocity of particles moving and reduces the particle velocity.

To study the radiation-drag effect, we have calculated a series of models, which are summarized in Table 1. For all the models, we set $M_{\rm BH}={10^8} M_{\odot}$,and $f_{\rm disk}=0.9$ (or $f_{X}$=0.1), respectively. Other parameters are listed in Table 1, where Col. (2) gives the bolometric luminosity, Col. (3) gives the initial vertical velocity ($v_{0}$), Col. (4) gives the initial particle number density ($n_{0}$), Col. (5) gives the fastest velocity of particles along streamlines, Cols. (6) and (7) also give the mass outflow rate and kinetic power of the escape particles, and Col. (8) gives the range of the velocity azimuth angles (relative to the disk surface) of the escape particles at the outer boundary, respectively. For comparing the properties of radiation-dragged particle winds with that of non-dragged particle winds, all the models are classified into A-group and B-group models, respectively. For the A-group models (models A0--A17) without the radiation-drag effects, we remove $-E_0 v^i-\sum v_j P^{ij}_0$  from equation (7), while all the terms are included in equation (7) for the B-group models (models B0--B17) with the radiation-drag effects.

Figure 1 shows the trajectories of winds in models A3 and B3 (panels (A) and (B)) and models A13 and B13 (panels (C) and (D)). Models A13 and B13 have higher initial velocity than models A3 and B3. In the top panel, we can see that the formation region of winds is different in models A3 and B3. The escape winds of model B3 are generated in a smaller region than that of model A3. This implies that the radiation-drag effect reduces the radiation force of driving particles and then more particles cannot reach the escape velocity. According to equations (A1) and (A3) in Appendix A, the force multiplier of line force is positively correlated with the acceleration of particles and is inversely correlated with the speed of particles. When the radiation flux in the comoving frame becomes weak due to the radiation-drag effects, the acceleration of particles becomes small due to the reduction of radiation force and then the force multiplier of line force further becomes small so that the line force is weakened. Therefore, the radiation-drag effect can weaken the line force by directly reducing the radiation flux in the comoving frame and weakening the line force multiplier. That is the reason the escape winds in model B3 form within a smaller region and the opening angle of winds becomes smaller, compared with the case of model A3. In panel (D), we can see that the wind particles of model B13 cannot escape to infinity and eventually they fall back on the disk surface. In models A13 and B13, we set a higher initial velocity than that in models A3 and B3. The higher initial velocity weakens the line force multiplier and therefore the line force becomes weaker in models A13 and B13. Comparing models A3 and A13, the initial velocity in model A13 is one order of magnitude higher than that in model A3 (see Table 1), but the maximum velocity of winds is slightly three quarters of that in model A3, while the mass outflow rate exceeds seven times that in model A3 and kinetic power are almost three times that in model A3. In model A13, the line force becomes weak, which causes that the maximum velocity of winds becomes slower than that in model A3. However, the higher initial velocity in model A13 makes more particles reach the escape velocity and therefore the mass outflow rate and kinetic power of winds are higher than that in model A3. Furthermore, due to the radiation-drag effects, the radiation force in model B13 is too weak to accelerate particles to the escape velocity.

The radiation drag term in equation (7) includes a component ($E_0v^i$) related to the radiation energy density and a component ($v_j P_0^{ij}$) related to the radiation pressure tensor. Figure 2 shows not only the ratio ($\frac{E_0v^i}{v_j P_0^{ij}}$) between the two components along a streamline (i.e. the thick line in panel (B) of figure 1), but also the ratio of the radiation-drag term to the radiation flux ($F^{i}_{0}$) in the rest frame. The radiation-drag term in the radial and vertical direction is less than 15\% of the radiation flux in the rest frame, and the $E_0v^i$ component is stronger than the $v_j P_0^{ij}$ component on most areas of the streamline while the $v_j P_0^{ij}$ component should not be ignored. When the Doppler effect of rotating of the thin disk is ignored, the radiation flux ($F^{\phi}_0$) in the azimuthal direction equals 0 since the radiation field is axisymmetric. When the Doppler effect of rotating is included, $F^{\phi}_0$ is also a small quantity. In the radiation-drag term, the $E_0v^{\phi}$ component is dominant. When the $\phi$-direction radiation drag term is ignored, the angular momentum of particles is conserved along their streamlines. When the $\phi$-direction radiation drag term is included in dynamics equations, the angular momentum of particles is slightly reduced along their streamlines and then the centrifugal force of particles is weakened slightly, compared with the case of angular momentum conservation. However, the $\phi$-direction radiation drag term is still negligible in the dynamics of particles.

\subsection{Dependence on related parameters: $L_{\rm bol}$, $v_{0}$, and $n_{0}$}

In this section, we test the effects of three parameters (such as $L_{\rm bol}$, $v_0$, and $n_0$) on the wind properties. For models A0--A9 and models B0--B9 in Table 1, $L_{\rm bol}$ varies from 0.01 to 0.8 $L_{\rm Edd}$ while $v_0$ and $n_0$ are set to be $1\times10^5$ m/s and $5 \times 10^7$ cm$^{-3}$, respectively. Figure 3 shows the influences of $L_{\rm bol}$ on the maximum speed along streamlines, the mass outflow rate, and the kinetic power of winds. When a particle reaches the escape velocity, we take into account the contribution of the particle to the mass outflow rate and the kinetic power of winds. Due to that the mass of wind particles is constant along streamlines, the mass outflow rate and the kinetic power can be given by
\begin{equation}
\dot M_{\rm out}=4\pi\sum_i{\rho_{0,i}v_{0,i}R_{0,i}\Delta L_{0,i}},
\end{equation}
and
\begin{equation}
P_{\rm kin}=2\pi\sum_i{\rho_{0,i}v_{0,i}R_{0,i}\Delta L_{0,i}} v_{i}^2,
\end{equation}
respectively, where $i$ represents the $i$-th escape particle, $v_{i}$ is the particle velocity at the outer boundary, and $\rho$ ($\equiv m_{\rm p} n$)is the density of particles, respectively. The emitting angles ($\theta_{\rm out}$) of winds are given by
\begin{equation}
 \theta_{\rm out}=Arctan(\frac{v^z_{\rm out}}{v^R_{\rm out}}),
\end{equation}
where $v^z_{\rm out}$ and $v^R_{\rm out}$ are the vertical and radial velocities of escape particle at the outer boundary, respectively.

Panel (A) of figure 3 shows the dependence of the maximum speed along streamlines on the bolometric luminosity. In the A-group models without the radiation-drag effects, winds are very weak when $L_{\rm bol}<$0.05 $L_{\rm Edd}$ and winds cannot be generated when $L_{\rm bol}<$0.01 $L_{\rm Edd}$. In the B-group models including the radiation-drag effects, winds cannot be generated when $L_{\rm bol}<$0.05 $L_{\rm Edd}$. For models A2 and B2, whose luminosity is 0.1 $L_{\rm Edd}$, the maximum speed along streamlines can reach 0.228 $c$ and 0.08 $c$, respectively. In the A-group models, as $L_{\rm bol}$ increases from 0.01 to 0.3 $L_{\rm Edd}$, the maximum speed increases rapidly, while when $L_{\rm bol}>$ 0.3 $L_{\rm Edd}$ the maximum speed increase slightly with the increase of bolometric luminosity. In the B-group models, the maximum speed is significantly lower than that in the A-group models. According to equation (12), the mass outflow rate is determined by the disk surface area emitting the escaped particles. In the A-group models, the emission region of winds becomes wider and farther away from the central BH with the increase of $L_{\rm bol}$. In the B-group models, the emission region of winds almost does not change when $L_{\rm bol}>$0.6 $L_{\rm Edd}$. Comparing the A-group and B-group models, the emission region of winds in the B-group models is narrower than that of A-group models, so that the mass outflow rate in the B-group models is significantly lower than that in the A-group models, as shown in Panel (B).

Panel (C) shows the dependence of the wind power ($L_{\rm kin}$) on the bolometric luminosity. According to equation (13), the wind power depends on the areas where winds are generated and the velocity at the outer boundary. The higher the disk luminosity, the stronger the winds, as shown in panel (C). Due to the radiation-drag effects, the winds in the B-group models are significantly weaker than that in the A-group models. Panel (D) shows that the radiation-drag effects make winds closer to the disk surface when $L_{\rm bol}>$0.3 $L_{\rm Edd}$, while when $L_{\rm bol}<$0.3 $L_{\rm Edd}$ the opening angles of winds in the B-group models become narrower.

Figure 4 shows the influences of the initial velocity ($v_{0}$) of particles on the properties of winds. For models A3 and A10--A13 and models B3 and B10--B13, that shown in Figure 4, $L_{\rm bol}$ and $n_0$ are set to be 0.2 $L_{\rm Edd}$ and $5 \times 10^7$ $\text{cm}^{-3}$, respectively, while $v_0$ varies from $10^4$ to $10^6$ m/s, as given in Table 1. Panel (A) shows that the maximum speed along streamlines is significantly reduced due to the increase of the initial velocity. For model A10, the maximum speed reaches 0.384 $c$, while in model A13 the maximum speed is 0.170 $c$. The reason is that line force is weaken due to the increase of particle speed according to equations (A1) and (A3) in Appendix A. Therefore, the wind particles can not be accelerated to higher velocity, when the initial velocity increases. When the radiation-drag effects are considered, the maximum speed is reduced by $50 \% \sim 100\% $, because the decrease of acceleration is caused by radiation drag. For models A3 and A10--A13, the mass outflow rate and the kinetic power increase rapidly with the increase of the initial velocity, because the increase of initial velocity causes that more particles are emitted per unit time from the disk surface. However, when the radiation-drag effects are considered, the kinetic power in the B-group models decreases with the increase of $v_{0}$ when $v_{0}>5\times10^4$ m/s, and the mass outflow rate increases slightly with the increase of $v_{0}$. For model B13 ($v_{0}=10^6$ m/s), the mass outflow rate and the kinetic power decrease to 0, because the particles cannot escape to the infinity because line force is significantly weaken. Compared with the A-group models, more particles in the B-group models are not accelerated to the escape velocity to form winds. As the initial velocity increases, the opening angles of winds decrease, as shown in Panel (D).

Figure 5 shows the influences of the particle number density ($n_{0}$) of wind particles on the properties of winds. For models A2 and A14--A17 and models B2 and B14--B17, $L_{\rm bol}$ and $v_0$ are set to be 0.2 $L_{\rm Edd}$  and $10^5$ m/s, respectively, while $n_0$ varies from $10^7$ to $10^9 \text{cm}^{-3}$, as given in Table 1. The maximum speed along streamlines and the kinetic power of winds increase with the number density of streamlines, as shown in Panel (A) and (C). Panel (B) shows that the mass loss rate increases with the number density when $n_{0}<5 \times 10^7 \text{cm}^{-3}$, keeps almost constant when $n_{0}\sim 5 \times 10^7$--$1 \times 10^8 \text{cm}^{-3}$, and then decreases when $n_{0}>10^8 \text{cm}^{-3}$. This reason is as follows. As pointed out in previous discussion, the mass loss rate depends on area of the disk surface from which the escape winds are emitted. The wider the area, the higher the mass loss rate. Higher density is helpful to shield the X-ray photons from the central corona and then the ionization degree of gases becomes low, so that line force becomes stronger in driving particles. When the gases become dense, the region from which the escape particles are emitted extends inward and becomes wide. However, when the gases become very dense, the UV photons produced in the inner region of disk are also shielded to some extent, which can weaken line force. The region from which the escape particles are emitted becomes narrow to some extent. In general, with the increase of number density, the particles are accelerated by line force to higher velocity and then stronger winds are generated. As the particle number density increases, the azimuth angles (relative the disk surface) of winds increases, as shown in Panel (D).

\section{Conclusions and Discussions}

In AGNs, when outflows/winds with mildly relativistic velocities move in the radiation field produced by the thin disk around a BH, the outflows are affected by the radiation-drag force. In this paper, we study the effects of the radiation-drag force on the properties of the line-force-driven winds, based on a non-hydrodynamic approach. In the approach, we neglect the gas pressure and take gases as particles. Compared with the case without the radiation-drag force, when the radiation-drag force is included in dynamics, the line-force-driven winds are significantly weakened. For example, the maximum speed of winds, the mass outflow rate, and the kinetic power of winds are significantly weakened. The details of our conclusions are summarized as follows:

1) The radiation-drag effects increase the lower limit of the bolometric luminosity that can drive winds. In A-group models without the radiation-drag effects, winds become very weak when $L_{\rm bol}<$0.05 $L_{\rm Edd}$. In the B-group models with the radiation-drag effects, the lower limit of the bolometric luminosity becomes 0.1 $L_{\rm Edd}$.

2) The radiation-drag force narrows the area where the winds are generated and then increases the area where the failed winds are formed.

3) The radiation-drag force makes the maximum speed of winds $\sim$60\%-70\% lower than that in the case without the radiation-drag force. In models A1--A9, the maximum speed of winds can reach $\sim$0.17--0.36 $c$, while in models B0--B9 the maximum speed can reach $\sim$0.11--0.15 $c$. When the density of winds increases, the line force is strengthened, and then the wind velocities increase.

The observation of the Fe XXV and Fe XXVI lines implies that UFOs move at $\sim$0.03--0.3 $c$, with a mean value of $\sim$0.14 $c$ (Tombesi et al. 2012). When the radiation-drag effects are considered, the line-force-driven winds are closer to the velocity range of UFOs. However, UFOs are highly ionized. The line-force-driven winds are low-ionization. Therefore, our results donot seem to be employed to explain the properties of UFOs. However, previous numerical simulations have implied that the line-force-driven winds may become highly-ionized winds (i.e. UFOs) when the winds are exposed to the X-ray photons (Normua \& Ohsuga 2017; Yang et al. 2021; Yang 2021). The density of the inner shielding gases varies with time. When the inner shielding gases become dense, the line force is efficient. When the inner shielding gases become thin, the line-force-driven winds become highly ionized. Therefore, the winds are intermittently accelerated by the line force to the mildly relativistic velocity. The time evolution of winds is not studied in present work. In the future, we will implement numerical simulations to examine the effects of the radiation-drag force on the line-force-driven winds. Another possibility is the effect of geometry. When winds are relatively close to the disk surface, X-ray photons are shielded by the inner dense gases above the thin disk, the ionization degree of winds is low, and then the winds are accelerated by the line force to the mildly relativistic velocity. When the winds leave the shielding area, they are ionized by X-rays photons and then form the observed UFOs.

There are some approximations in our calculations. For example, we ignore the frequency dependence of the line-force driving. It is often thought that the 200--3200$\AA$ UV radiation mainly contributes to the line-force driving (Proga \& Kallman 2004). In our calculations, there are three factors that may affect the strength of the line force. 1) In equation (7), the radiative flux (F$^i_0$), the radiation energy density (E$_0$), and the radiation stress tensor (P$^ij_0$) in the rest frame are the quantities integrated by the frequency throughout the entire range. This causes an overestimation of the $200-3200\AA$ UV radiation. For the disk of $L_{\rm disk}=0.6L_{Edd}$, the 200--3200$\AA$ UV luminosity is overestimated by $\sim$14\%. 2) The Doppler effects caused by the rotation of the accretion disk can affect the spectral shape and intensity measured in the rest frame. In our calculations, equation (B2) has taken into account the influence of Doppler effects on the intensity on the order of $v/c$. The Doppler shift of spectral shape can affect the integrated intensity over the 200--3200$\AA$ UV bands in the rest frame, which is ignored in our calculations. In the rest frame, when the Doppler shift of spectral shape is not included, the 200--3200$\AA$ UV flux in the rest frame is overestimated, compared to the case that includes the Doppler effects caused by disk rotation. For example, for the disk of $L_{\rm disk}=0.6L_{Edd}$, the 200--3200$\AA$ UV flux in the rest frame is overestimated by $\sim$5\% at most. 3) The Doppler effects caused by the motion of particles can further affect the spectral shape and intensity measured in the comoving frame with the moving particles, which changes the intensity integrated over the 200--3200$\AA$ UV bands in the comoving frame. Equation (7) has taken into account the correction of the Doppler effects on the radiative flux in the comoving frame on the order of $v/c$. For most B-group models, the maximum speed of particles is about $\sim$0.15$c$ and the particles move at $<0.15 c$ over most regions along the streamlines. The first-order accuracy of $v/c$ is acceptable in our calculations. Equation (7) ignores the Doppler shift of the spectral shape in the comoving frame. For simply estimating the correction, we set the disk to be the radiation source as a point and assume that a particle moves away from the radiation source at 0.15$c$. Then, when only the Doppler shift of the spectral shape is taken into account, the intensity integrated over the 200--3200$\AA$ UV bands in the rest frame is close to the integrated intensity in the comoving frame. Therefore, our calculations overestimate the 200--3200$\AA$ UV radiation of the line-force driving. However, it is acceptable that the frequency dependence of the line-force driving is ignored in our calculations.

\section{ACKNOWLEDGMENTS}
This work is supported in part by the Natural Science Foundation of China (grant 11973018 and 12147102). D. Bu is supported by the Natural Science Foundation of China (grant 12173065).

\section{DATA AVAILABILITY}
The data underlying this article will be shared on reasonable request to the corresponding author.

\begin{appendix}

\section{the line-force multiplier $\mathcal{M}$}
Based on the Sobolev approximation, the line-force multiplier ($\mathcal{M}$) is a function of the ionization parameter ($\xi$) and the local optical depth parameter ($t$) (Rybicki \& Hummer 1978). the line-force multiplier is given by
\begin{equation}
\mathcal{M}(t,\xi)=k(\xi)t^{-0.6}[\frac{(1+t\eta_{\rm max}(\xi))^{0.4}-1}{(t\eta_{\rm max}(\xi))^{0.4}}],
\end{equation}
where $k(\xi)$ and $\eta_{\rm max}(\xi)$ are a parameter proportional to the total number of lines and a parameter determining the maximum value of $\mathcal{M}$, respectively. The two parameters are a function of the ionization parameter. The local optical depth parameter ($t$) can be obtained by
\begin{equation}
t=\sigma_Tnv_{\rm th}|\frac{dv_{\textit{l}}}{d\textit{l}}|^{-1},
\end{equation}
where $\frac{dv_{\textit{l}}}{d\textit{l}}$ is the velocity gradient along the light of sight, $v_{\rm th}$ is the thermal velocity, and $\sigma_T$ is the Thomson cross section, respectively. For simplicity, the velocity gradient is written as
\begin{equation}
\frac{dv_{\textit{l}}}{d\textit{l}}=\frac{dv}{dt}\frac{dt}{d\textit{l}}=\frac{a}{v}=\frac{\sqrt{a_{\textit{\tiny{R}}}^2+a_z^2}}{\sqrt{v_{\textit{\tiny{R}}}^2+v_z^2}}.
\end{equation}
We set the gas thermal velocity $v_{\rm th}=20$ $\rm km \rm s^{-1}$, which corresponds to the gas temperature of $2.5 \times 10^4$ K (Stevens \& Kallman 1990; Proga 2007). For $k(\xi)$ and $\eta_{\rm max}(\xi)$, we use the data obtained from Stevens and Kallman (1990) to obtain the interpolation functions.

\section{the calculation of the radiation field in the rest frame}

\begin{figure}
 \scalebox{0.4}[0.4]{\includegraphics{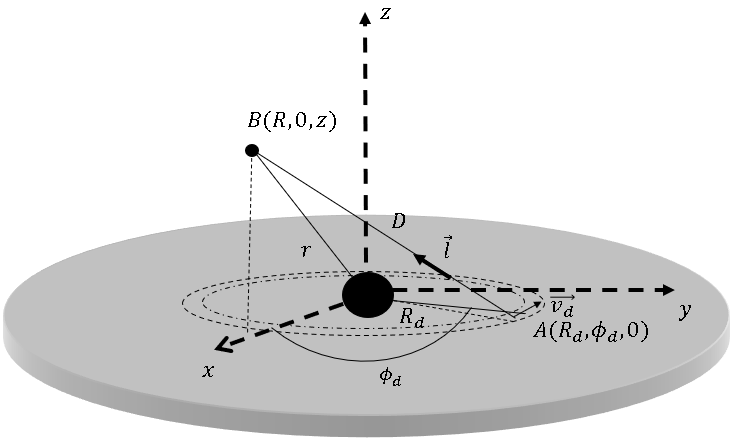}}\centering \caption{Schematic diagram for the calculation of the radiation field in the rest frame.}\label{fig:3}
\end{figure}

Figure B1 is a schematic diagram to display the calculation of the radiation field. In a cylindrical coordinate system ($R$, $\phi$, $z$), the position of an observer (B point) is given by $(R,0,z)$ while the position of an element (A point) on the disk is given by ($R_d,\phi_d,0$).

According to a theory of the standard thin disk, the local isotropic intensity on the disk surface of A is written as
\begin{equation}
I_d=\frac{3GM_{\rm BH}\dot{M}}{8 \pi^2 R_{\rm d}^3}(1-\sqrt{\frac{R_{\rm in}}{R_{\rm d}}}),
\end{equation}
where $\dot{M}$ is the mass accretion rate, $R_{\rm in}$ ($R_{\rm in}=3$ $r_{\rm s}$) the inner radius of the disk, and $R_{\rm d}$ the disk radius. The accretion disk rotates with the Keplerian velocity of $v_{K}=\sqrt{{GM_{BH}}/{R_{\rm d}}}$ and then the relativistic Doppler effect changes the radiation field in the rest frame standing at infinity. The radiation intensity measured by an observer rotating with the disk ($I_{\rm d}$) can be transformed to that measured by the rest observer ($I_0$) as follows:
\begin{equation}
I_0=\frac{I_{\rm d}}{(1-\frac{\overrightarrow{\textit{v}_{\rm d}}\cdot\overrightarrow{\textit{l}}}{c})^4},
\end{equation}
where $\overrightarrow{\textit{l}}$ is the direction-cosine vector pointing toward the rest observer from the location of the disk radiation and $\overrightarrow{\textit{v}_{\rm d}}$ is the disk velocity. In Cartesian coordinates, we can get
\begin{equation}
\overrightarrow{\textit{l}}=(\frac{R-R_{\rm d}{\rm cos}(\phi_{\rm d})}{D},\frac{-R_{\rm d}{\rm sin}(\phi_{\rm d})}{D},\frac{z}{D}),
\end{equation}
and
\begin{equation}
\overrightarrow{\textit{v}_{\rm d}}=(-v_{\rm K} \sin(\phi_{\rm d}),v_{\rm K}\cos(\phi_{\rm d}),0),
\end{equation}
where $D=(R^2+z^2+R_d^2-2RR_{\rm d}\cos(\phi_{\rm d}))^{\frac{1}{2}}$ is the distance from the rest observer to the element A on the disk surface, as shown in figure B1.

For the element A, its area $dS=R_{\rm d}d\phi_{\rm d}dR_{\rm d}$. The solid angle subtended by the element A towards the observer standing at the B point is $d\Omega=\frac{z dS}{D^3}$. The $E_{0}$, $F^{i}_{0}$, and $P^{ij}_{0}$ measured in the rest frame are expressed as (Tajima $\&$ Fukue 1998)
\begin{equation}\begin{split}
E_0=\frac{1}{c}\int I_0  d\Omega=\frac{3GM\dot{M}}{8c\pi^2r_{\rm s}^3}\int_{3}^{1500}\int_0^{2\pi}\frac{z}{R_{\rm d}^{2}}(1-\sqrt{\frac{3}{R_{\rm d}}})\\
\times\frac{1}{D^{3}(1-\frac{\overrightarrow{\textit{v}_{\rm d}}\cdot\overrightarrow{\textit{l}}}{c})^4}d\phi_{\rm d} dR_{\rm d},
\end{split}\end{equation}
\begin{equation}\begin{split}
F^{i}_{0}=\int I_0l^i d\Omega=\frac{3GM\dot{M}}{8\pi^2r_{\rm s}^3}\int_{3}^{1500}\int_0^{2\pi}\frac{z}{R_{\rm d}^{2}}(1-\sqrt{\frac{3}{R_{\rm d}}})\\
\times\frac{l^i}{D^{3}(1-\frac{\overrightarrow{\textit{v}_{\rm d}}\cdot\overrightarrow{\textit{l}}}{c})^4}d\phi_{\rm d} dR_{\rm d},
\end{split}\end{equation}
\begin{equation}\begin{split}
P^{ij}_{0}=\frac{1}{c} \int I_0l^il^j d\Omega=\frac{3GM\dot{M}}{8c\pi^2r_{\rm s}^3}\int_{3}^{1500}\int_0^{2\pi}\frac{z}{R_{\rm d}^{2}}(1-\sqrt{\frac{3}{R_{\rm d}}})\\
\times\frac{l^il^j}{D^{3}(1-\frac{\overrightarrow{\textit{v}_{\rm d}}\cdot\overrightarrow{\textit{l}}}{c})^4}d\phi_{\rm d} dR_{\rm d}.
\end{split}\end{equation}

\end{appendix}
\end{document}